\documentclass{article}
\usepackage{spconf,amsmath,graphicx}
\usepackage{booktabs,url}

\newcommand{\Fig}[1]{\textbf{Fig. \ref{fig:#1}}} 

\newcommand{\Eq}[1]{Eq. (\ref{eq:#1})} 
\newcommand{\Eqs}[1]{Eqs. (\ref{eq:#1})}
\newcommand{\Eqss}[1]{(\ref{eq:#1})}
\newcommand{\Sec}[1]{\textbf{Section \ref{sec:#1}}} 

\newcommand{\Norm}[1]{ {\mathcal N}\left( #1 \right) } 
 
\renewcommand{\Vec}[1]{\textrm{\boldmath $#1$}} 

\newcommand{\pt}[1]{\left(#1\right)} 
\newcommand{\br}[1]{\left[#1\right]} 
\newcommand{\x}{ \Vec{x} } 
\newcommand{\z}{ \Vec{z} } 
\newcommand{\cc}{ \Vec{c} } 
\newcommand{\hatx}{ \Vec{\hat x} } 

\newcommand{\drawfig}[4]{ 
  \begin{figure}[#1]
  \centering \vspace{-0mm}
  \includegraphics[width=#2,clip]{#3.pdf} \vspace{-3mm} 
  \caption{#4} \vspace{-4mm}
  \label{fig:#3}
  \end{figure}
}
\newcommand{\drawfigdown}[4]{ 
  \begin{figure}[#1]
  \centering \vspace{4mm}
  \includegraphics[width=#2,clip]{#3.pdf} \vspace{-3mm} 
  \caption{#4} \vspace{-4mm}
  \label{fig:#3}
  \end{figure}
}

\title{HumanACGAN: conditional generative adversarial network\\with human-based auxiliary classifier\\and its evaluation in phoneme perception}

\makeatletter
\def\name#1{\gdef\@name{#1\\}}
\makeatother
\name{{\em Yota Ueda$^{1}$, Kazuki Fujii$^{2}$, Yuki Saito$^{1}$, Shinnosuke Takamichi$^{1}$, Yukino Baba$^{3}$ and Hiroshi Saruwatari$^{1}$}}
\address{
    $^1$ Graduate School of Information Science and Technology, The University of Tokyo, Japan. \\
    $^2$ National Institute of Technology, Tokuyama College, Japan. \\
    $^3$ Faculty of Engineering, Information and Systems, University of Tsukuba, Japan. \\
}

\begin{document}
\ninept
\maketitle

\setlength{\abovedisplayskip}{3pt} 
\setlength{\belowdisplayskip}{3pt} 
\setlength\textfloatsep{22pt} 

\allowdisplaybreaks

\begin{abstract} \vspace{-1mm}
    We propose a conditional generative adversarial network (GAN) incorporating humans' perceptual evaluations. A deep neural network (DNN)-based generator of a GAN can represent a real-data distribution accurately but can never represent a human-acceptable distribution, which are ranges of data in which humans accept the naturalness regardless of whether the data are real or not. A HumanGAN was proposed to model the human-acceptable distribution. A DNN-based generator is trained using a human-based discriminator, i.e., humans' perceptual evaluations, instead of the GAN's DNN-based discriminator. However, the HumanGAN cannot represent conditional distributions. This paper proposes the HumanACGAN, a theoretical extension of the HumanGAN, to deal with conditional human-acceptable distributions. Our HumanACGAN trains a DNN-based conditional generator by regarding humans as not only a discriminator but also an auxiliary classifier. The generator is trained by deceiving the human-based discriminator that scores the unconditioned naturalness and the human-based classifier that scores the class-conditioned perceptual acceptability. The training can be executed using the backpropagation algorithm involving humans' perceptual evaluations. Our experimental results in phoneme perception demonstrate that our HumanACGAN can successfully train this conditional generator.
\end{abstract}


\vspace{-2mm}
\section{Introduction} \vspace{-2mm}
    Deep generative models of machine learning have contributed to media research~\cite{goodfellow14gan,kingma2013vae,dinh2014nice}. A generative adversarial network (GAN)~\cite{goodfellow14gan} is one of the strongest generative models. It has been applied in speech modeling~\cite{saito18advss,hono2019singing}. The GAN consists of a set of deep neural networks (DNNs), a generator, and a discriminator. The generator is trained to deceive the discriminator, and the discriminator is trained to distinguish between real and generated data. After iterating them, the trained generator represents a real-data distribution and can randomly generate data that follows the real-data distribution. 
    
    The GAN cannot represent an outer side of the real-data distribution. However, humans can accept out-sided media as natural. In speech perception, humans can recognize a voice as a human's even though the voice is out of the real-data distribution (i.e., an actual humans' voice). For example, that range contains synthesized voices or processed voices. In this study, we call this data range perceived by humans the \textit{human-acceptable distribution}. HumanGAN~\cite{fujii2020humangan} was proposed to model the human-acceptable distribution by using humans as the discriminator of the GAN. The top of ~\Fig{intro} shows the comparison of a GAN and HumanGAN. The HumanGAN regards humans as a black-boxed system that outputs a difference in posterior probabilities given generated data. The DNN-based generator is trained using the backpropagation algorithm, including the human-based discriminator. The trained generator can represent the human-acceptable distribution.
    
    However, the HumanGAN's generator cannot achieve more practical generative modeling such as text-to-speech synthesis~\cite{sagisaka88} and voice conversion~\cite{stylianou88} because it cannot represent conditional distributions. Because the GAN was extended to the conditional GAN~\cite{mirza2014conditional,odena2017conditional}, we expect that the HumanGAN can be extended to the conditional modeling. Namely, we train the HumanGAN's generator conditioned on the desired class label, and it represents the class-specific human-acceptable distribution as a result. This will contribute establishing a DNN-based framework to model the task-oriented perception by humans~\cite{chiu2020human,peterson2018capturing}.
        \drawfig{t}{0.98\linewidth}{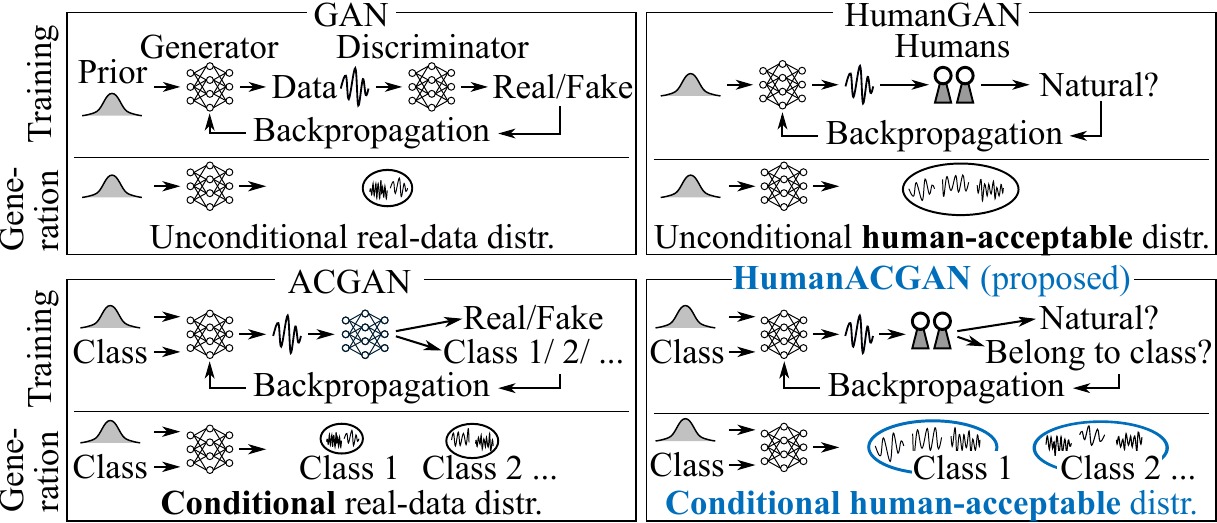}
        {Comparison of four GANs. We extend the HumanGAN to the conditional modeling in the same way the GAN was extended to the ACGAN. While an ACGAN trains a conditional generator with a DNN-based discriminator and an auxiliary classifier, the HumanACGAN trains one with a humans-based discriminator and an auxiliary classifier. The trained generator of the HumanACGAN can represent human-acceptable distributions conditioned by input class labels.} 
        
    In this paper, we propose the \textit{HumanACGAN}, aiming to train a DNN-based conditional generator using humans' perceptual evaluations. To this end, we extend the HumanGAN by introducing an auxiliary classifier GAN (ACGAN)~\cite{odena2017conditional}: class-conditional expansion of a GAN. \Fig{intro} shows a comparison of four GANs: a GAN, an ACGAN, a HumanGAN, and our HumanACGAN. The ACGAN uses a DNN-based auxiliary classifier to train a conditional DNN-based generator in addition to a discriminator of the GAN. Our HumanACGAN replaces both the DNN-based discriminator and auxiliary classifier with humans. The HumanACGAN's generator is trained using human-perception-based discrimination and classification. This training is operated using an expansion of the backpropagation-based algorithm incorporating humans' perceptual evaluations as a discriminator and an auxiliary classifier. We evaluated the HumanACGAN in phoneme perception, a task to train a DNN-based generator that represents phoneme-specific human-acceptable distributions. The experimental results show that 1) the phoneme-conditioned human-acceptable distributions are wider than the real-data ones and that 2) the HumanACGAN can successfully train a generator that represents conditional human-acceptable distributions.

\vspace{-1mm}
\section{Related Works} \vspace{-2mm}
    \subsection{ACGAN} \vspace{-1mm}
        The ACGAN~\cite{odena2017conditional} trains a DNN-based conditional generator that represents real-data distributions conditioned on class labels. To achieve this, the ACGAN uses a DNN-based auxiliary classifier in addition to a DNN-based discriminator of the GAN. The generator $G\pt{\cdot}$ transforms a prior noise $\z = \br{\z_1, \cdots, \z_n, \cdots, \z_N}$ into data $\hatx = \br{\hatx_1, \cdots, \hatx_n, \cdots, \hatx_N}$, conditioned on class labels $\cc = \br{c_1, \cdots, c_n, \cdots, c_N}$, i.e., $\hatx_n = G\pt{\z_n,\cc_n}$. $N$ denotes the number of data. The prior noise follows a known probability distribution, e.g., a uniform distribution $U\pt{0,1}$. Here, let real data be $\x = \br{\x_1, \cdots, \x_n, \cdots, \x_N}$. The real data also have the same corresponding class label $\cc$. A discriminator $D_{\mathrm{S}}\pt{\cdot}$ and a classifier $D_{\mathrm{C}}\pt{\cdot}$ are used for the generator training. The $D_{\mathrm{S}}\pt{\cdot}$ takes $\x_n$ or $\hatx_n$ as an input and outputs a posterior probability that the input is real data. The $D_{\mathrm{C}}\pt{\cdot}$ takes an $\x_n$ or $\hatx_n$ as an input and outputs a posterior probability that the input source belongs to each class. Objective functions in training are formulated as
            \begin{align}
                \hspace{-2mm} L_{\mathrm{S}} &= \sum\limits_{n=1}^N \log D_{\mathrm{S}}\pt{\x_n} + \sum\limits_{n=1}^N \log \pt{1 - D_{\mathrm{S}}\pt{G\pt{\z_n,\cc_n}}}, \\
                \hspace{-2mm} L_{\mathrm{C}} &= \sum\limits_{n=1}^N \log D_{\mathrm{C}}\pt{\x_n,\cc_n} + \sum\limits_{n=1}^N \log \pt{D_{\mathrm{C}}\pt{G\pt{\z_n, \cc_n},\cc_n}}.
                \label{eq:gan_loss}
            \end{align}
        $L_{\mathrm{S}}$ is the objective function of the GAN, and $L_{\mathrm{C}}$ enables training the conditional generator. The generator is trained to maximize $-L_{\mathrm{S}} + \lambda L_{\mathrm{C}}$, where $\lambda$ is a hyperparameter. 

        The ACGAN trains the generator using real data, and the generator represents the real-data distribution conditioned on the data class. However, because the human-acceptable distribution is wider than the real-data distribution~\cite{fujii2020humangan}, the ACGAN cannot represent the full range of that distribution.

    \subsection{HumanGAN} \vspace{-1mm}
        The HumanGAN~\cite{fujii2020humangan} was proposed to represent the human-acceptable distribution in a wider manner than a real-data distribution. A DNN-based unconditional generator of the HumanGAN is trained using humans' perceptual evaluations instead of the DNN-based discriminator of the GAN. The generator $G\pt{\cdot}$ transforms prior noise $\z_n$ to data $\hatx_n$ unconditionally. A human-based discriminator $D_{\mathrm{S}}\pt{\cdot}$ is defined to deal with humans' perceptual evaluations. $D_{\mathrm{S}}\pt{\cdot}$ takes $\hatx_n$ as an input and outputs a posterior probability that the input is perceptually acceptable. The objective function is
            \begin{align}
                L_{\mathrm{S}} = \sum\limits_{n=1}^N D_{\mathrm{S}}\pt{G\pt{\z_n}}.
                \label{eq:humangan_loss}
            \end{align}
        The generator is trained to maximize $L_{\mathrm{S}}$. A model parameter $\theta$ of $G\pt{\cdot}$ is iteratively updated as $\theta \leftarrow \theta + \alpha \partial L_{\mathrm{S}}/{\partial \theta}$, where $\alpha$ is the learning coefficient, and $\partial L_{\mathrm{S}} / \partial \theta = \partial L_{\mathrm{S}} / \partial \hatx_n \cdot \partial \hatx_n / \partial \theta$.
        
        $\partial\hatx_n/\partial\theta$ can be estimated analytically, but $\partial L_{\mathrm{S}}/{\partial \hatx_n}$ cannot because the human-based discriminator $D_{\mathrm{S}}\pt{\cdot}$ is not differentiable. The HumanGAN uses the natural evolution strategy (NES)~\cite{ilyas18blackboxlimitedquery} algorithm to approximate the gradient. A small perturbation $\Delta \x_{n}^{(r)}$ randomly generated from a multivariate Gaussian distribution $\Norm{\Vec{0}, \sigma^2\Vec{I}}$, and it is added to a generated datum $\hatx_n$. $r$ is the perturbation index $\pt{1 \leq r \leq R}$. $\sigma$, $\Vec{I}$ are the standard deviation and the identity matrix, respectively. Next, a human observes two perturbed data $\{ \hatx_n + \Delta \x_{n}^{(r)}, \hatx_n - \Delta \x_{n}^{(r)} \}$ and evaluates the difference in their posterior probabilities of naturalness:
            \begin{align}
                \Delta D_{\mathrm{S}}\ (\hatx_n^{(r)}) \equiv D_{\mathrm{S}}\ (\hatx_n + \Delta \x_{n}^{(r)}) - D_{\mathrm{S}}\ (\hatx_n - \Delta \x_{n}^{(r)}).
                \label{eq:D_S}
            \end{align}
        $\Delta D_{\mathrm{S}}\ (\hatx_n^{(r)})$ ranges from $-1$ to $1$. For instance, a human will answer $\Delta D_{\mathrm{S}}\ (\hatx_n^{(r)})=1$ when he/she perceives that $\hatx_n + \Delta \x_{n}^{(r)}$ is substantially more acceptable than $\hatx_n - \Delta \x_{n}^{(r)}$. $\partial L_{\mathrm{S}} / \partial \hatx$ is approximated with~\cite{ilyas18blackboxlimitedquery}
            \begin{align}
                \frac{\partial L_{\mathrm{S}}}{\partial \hatx_n}&= \frac{1}{2\sigma^2 R} \sum\limits_{r=1}^{R} 
                \Delta D_{\mathrm{S}}\ \pt{\hatx_n^{(r)}} \cdot \Delta \x_{n}^{(r)}. 
                \label{eq:grad_S}
            \end{align}

\vspace{-1mm}
\section{HumanACGAN} \vspace{-2mm}
    \subsection{Training}
        \drawfig{t}{0.9\linewidth}{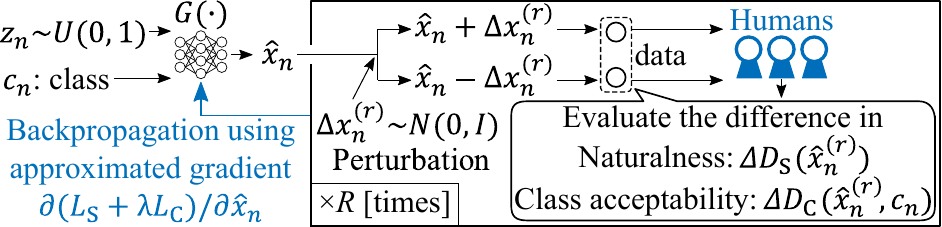}
        {Generator training process of proposed HumanACGAN. A human observes two perturbed data and evaluates their perceptual difference in two views: naturalness and class acceptability. They give global and class-specific gradients, respectively. Evaluations and perturbations are used for backpropagation to train the generator.}
        
        We propose the HumanACGAN. As shown in \Fig{achumangan}, the HumanACGAN has a conditional generator that represents class-specific human-acceptable distributions, while the HumanGAN only deals with one class. As well as the ACGAN, our HumanACGAN consists of a conditional generator, a discriminator, and an auxiliary classifier. The DNN-based generator $G\pt{\cdot}$ is the same as that of the ACGAN; it transforms prior noise $\z_n$ into data $\hatx_n$ conditioned on a class label $\cc_n$. The discriminator and the auxiliary classifier are redefined to incorporate humans' perceptual evaluations. The human-based discriminator $D_{\rm S}\pt{\cdot}$ is the same as that of the HumanGAN; it outputs a posterior probability of global (i.e., class-independent) naturalness. In addition, the human-based auxiliary classifier $D_{\mathrm C}\pt{\cdot}$ evaluates class acceptability, whether or not the data can belong to the class $\cc_n$. $D_{\mathrm C}\pt{\cdot}$ takes $\hatx_n$ generated from $G\pt{\cdot}$ and class label $\cc_n$ as inputs and outputs a posterior probability that the input is perceptually acceptable \textit{as the class}''. The objective functions are
            \begin{align}
                L_{\mathrm{S}} =& \sum\limits_{n=1}^N D_{\mathrm{S}}\pt{G\pt{\z_n,\cc_n}}
                \label{eq:L_S}, \\
                L_{\mathrm{C}} =& \sum\limits_{n=1}^N D_{\mathrm{C}}\pt{G\pt{\z_n,\cc_n},\cc_n}
                \label{eq:L_C}.
            \end{align}
        A model parameter $\theta$ of $G\pt{\cdot}$ is estimated by maximizing $L_{\mathrm{S}} + \lambda L_{\mathrm{C}}$. $\theta$ is iteratively updated as follows.
            \begin{align}
                \theta \leftarrow& \theta + \alpha \frac{\partial\pt{ L_{\mathrm{S}} + \lambda L_{\mathrm{C}}}}{\partial \theta}, \\
                \frac{\partial\pt{ L_{\mathrm{S}} + \lambda L_{\mathrm{C}}}}{\partial \theta} =& \pt{\frac{\partial L_{\mathrm{S}}}{\partial \hatx_n} + \lambda\frac{\partial L_{\mathrm{C}}}{\partial \hatx_n}} \cdot \frac{\partial\hatx_n}{\partial\theta}.
            \end{align}
        As well as the HumanGAN, $\partial\hatx_n/\partial\theta$ can be estimated analytically, but $\partial L_{\mathrm{S}}/{\partial \hatx_n}$ and $\partial L_{\mathrm{C}}/{\partial \hatx_n}$ cannot. We formulate gradient approximation using the NES algorithm. A human observes two perturbed data $\{ \hatx_n + \Delta \x_{n}^{(r)}, \hatx_n - \Delta \x_{n}^{(r)} \}$ and evaluates two kinds of difference in their posterior probabilities. The first is the same as the HumanGAN, i.e., the human evaluates ``to what degree the inputs are perceptually different in the view of naturalness'' and answers $\Delta D_{\mathrm{S}}\ (\hatx_n^{(r)})$ for approximating $\partial L_{\mathrm{S}}/{\partial \hatx_n}$ as in \Eq{D_S}. The second is for the class-specific question, ``to what degree the inputs are different in the view of class acceptability.'' The difference in the posterior probability $\Delta D_{\mathrm{C}}\ (\hatx_n^{(r)}, \cc_n)$ is defined as
            \begin{align}
                \Delta D_{\mathrm{C}}(\hatx_n^{(r)}, \cc_n) \equiv& D_{\mathrm{C}}\ \pt{\hatx_n + \Delta \x_{n}^{(r)}, \cc_n}\nonumber\\
                &- D_{\mathrm{C}}\ \pt{\hatx_n - \Delta \x_{n}^{(r)}, \cc_n}.
            \end{align}
        In the same way as the HumanGAN, $\partial L_{\mathrm S}/{\partial \hatx_n}$ can be approximated using $\Delta D_{\mathrm{S}}\ (\hatx_n^{(r)})$. Unlike $\Delta D_{\mathrm{S}}\pt{\cdot}$, $\Delta D_{\mathrm{C}}\pt{\cdot}$ becomes the class-specific difference. Namely, $\Delta D_{\mathrm{C}}\ (\hatx_n^{(r)}, \cc_n)$ will be 1 when the human perceives that $\hatx_n + \Delta \x_{n}^{(r)}$ is substantially more acceptable than $\hatx_n - \Delta \x_{n}^{(r)}$ as the presented class. $\Delta D_{\mathrm C}\ (\hatx_n^{(r)},\cc_n)$ ranges from $-1$ to $1$, and $\partial L_{\mathrm C} / \partial \hatx$ is approximated with
            \begin{align}
                \frac{\partial L_{\mathrm{C}}}{\partial \hatx_n}=
                \frac{1}{2\sigma^2 R} \sum\limits_{r=1}^{R} 
                \Delta D_{\rm C}\ \pt{\hatx_n^{(r)},\cc_n} \cdot \Delta \x_{n}^{(r)}. 
                \label{eq:grad_C}
            \end{align}
        Note that the HumanACGAN does not explicitly involve classification problems. In other words, it only needs to estimate the gradient of multiclass probability function (i.e., the softmax function used in an ACGAN) as the degree of class-specific acceptability. 
        
    \subsection{Limitations}
        The HumanGAN~\cite{fujii2020humangan} suffers from a mode collapse problem~\cite{goodfellow16gantutorial}, a gradient vanishing, and scalability to the data size and dimensionality. These problems still remain in our HumanACGAN. Therefore, our experiments followed the initialization way and data preprocessing of the paper~\cite{fujii2020humangan}. We first estimated histograms of the posterior probabilities and initialized $\theta$ by referring to the histograms. Also, we reduced the data dimensionality using principal component analysis (PCA).

\vspace{-1mm}
\section{Experimental evaluation} \vspace{-2mm}
This section describes an experimental test of the effectiveness of the HumanACGAN using Japanese phonemes as classes. 
    \subsection{Experimental setup} \label{sec:expcond}

        Two phonemes we used were /i/ and /e/ of Japanese vowels. We basically followed the experimental setup of the HumanGAN paper~\cite{fujii2020humangan}. The used data consisted of $199$ female speakers' utterances recorded in the JVPD~\cite{jvpd_corpus} corpus. Before extracting speech features, the speech waveforms were downsampled at $16$~kHz, and their powers were normalized. 513-dimensional log spectral envelopes, fundamental frequency (F0), and aperiodicities (APs) were extracted every $5$ ms from the speech waveforms using the WORLD vocoder~\cite{morise16world,morise16d4c}. We extracted the speech features of the vowels /i/ and /e/ using phoneme alignment obtained by Julius~\cite{lee01julius}. We applied PCA to the log spectral envelopes and used the first and second principal components. The two-dimensional principal components were normalized to have zero-mean and unit-variance. The speech waveforms to be evaluated were synthesized using features obtained from a DNN-based conditional generator in the following way. First, the first and second principal components were generated by the generator and de-normalized. For the other features, i.e., the F0 and the APs, we used the average of all speakers. These corresponded to the speech features of one frame ($5$ ms). Next, we copied the features for $200$ frames to make the perceptual evaluations easy and synthesized $1$ second of speech waveforms using the WORLD vocoder.
        
        \drawfig{t}{0.98\linewidth}{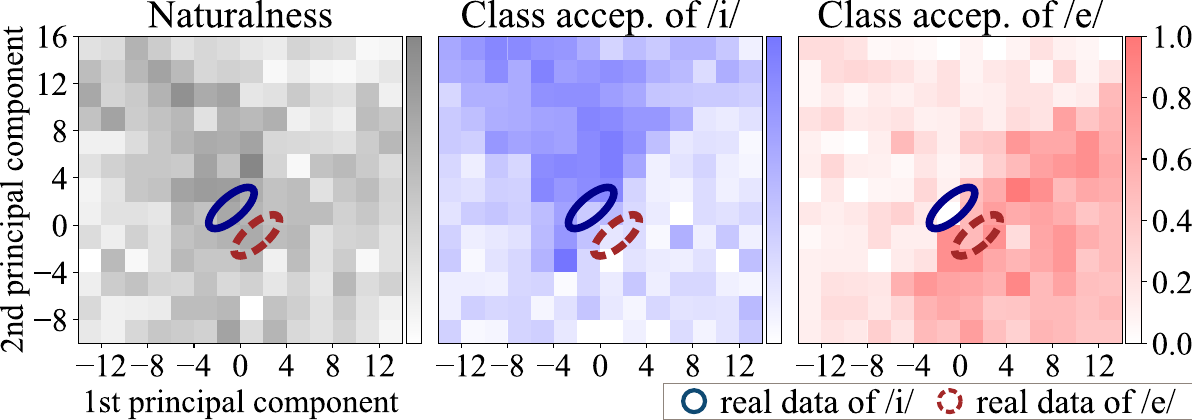}
        {Color maps representing posterior probabilities of naturalness and class acceptability (``accep.'') of /i/ and /e/.}
        
        \drawfigdown{t}{0.96\linewidth}{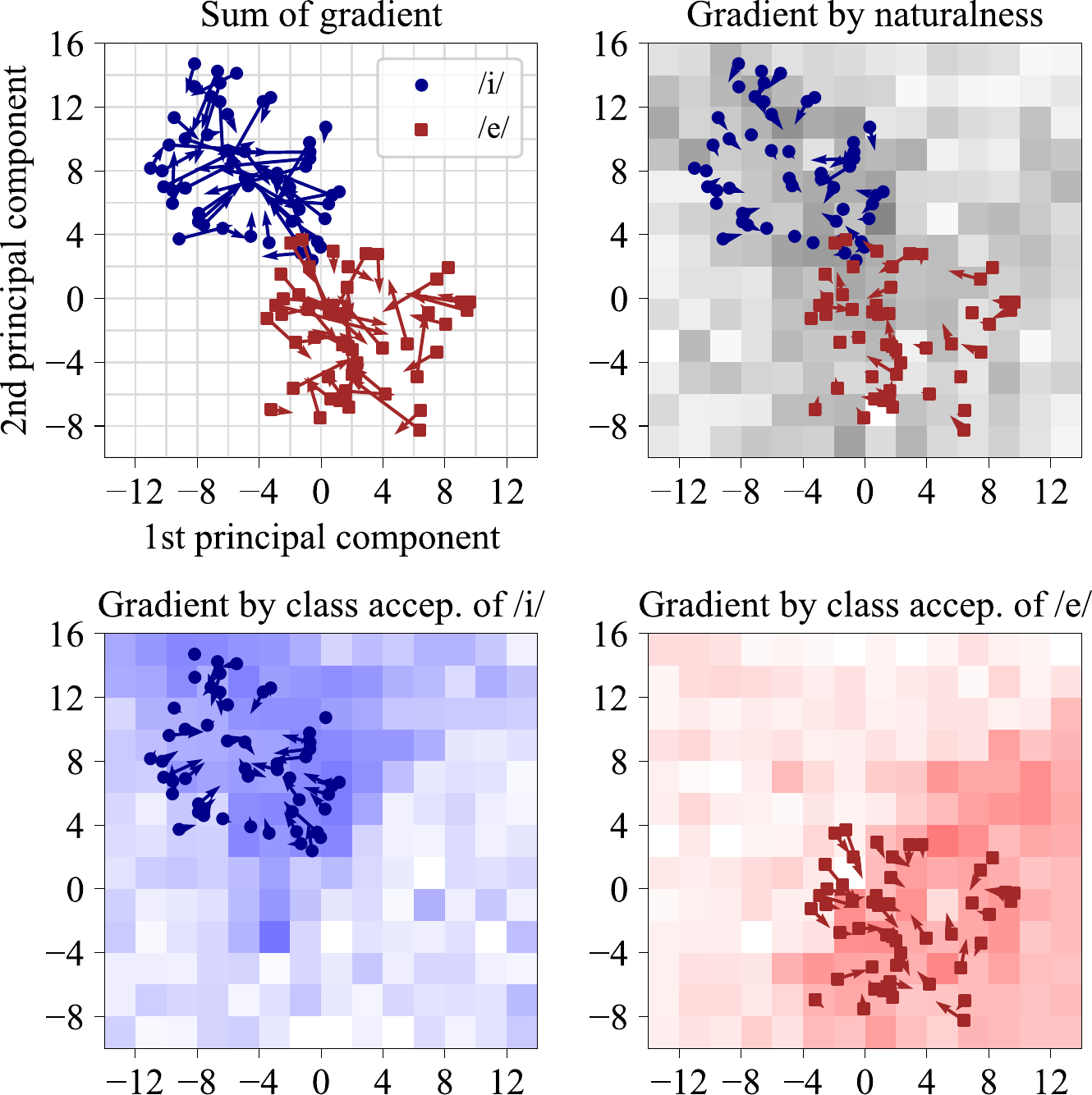}{Generated data (points) and gradient (arrows) estimated by our algorithm. ``accep.'' denotes acceptability. The gradients of the upper left figure are the weighted sum of the other three gradients. We can see that the gradients for naturalness and class acceptability are respectively pointing to darker (i.e., higher posterior) zones.}

        \drawfig{t}{0.9\linewidth}{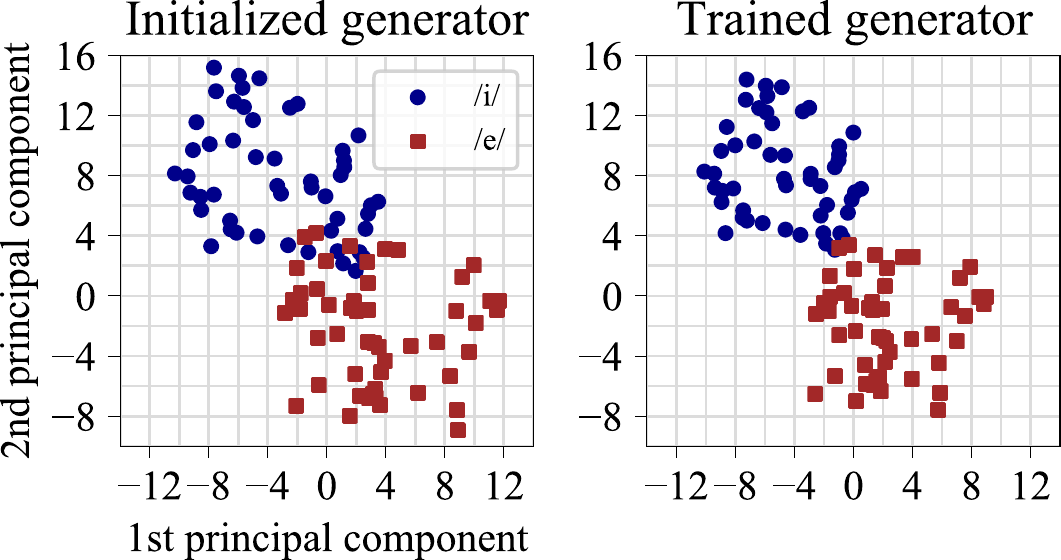}{Data generated from the initialized or trained generator. The colors of the data points correspond to those of the color maps for class acceptability in \Fig{apos}. Based on posterior probabilities shown in \Fig{apos}, we can say that the training makes the data move to have higher posterior probabilities.}
        
        \drawfigdown{t}{0.98\linewidth}{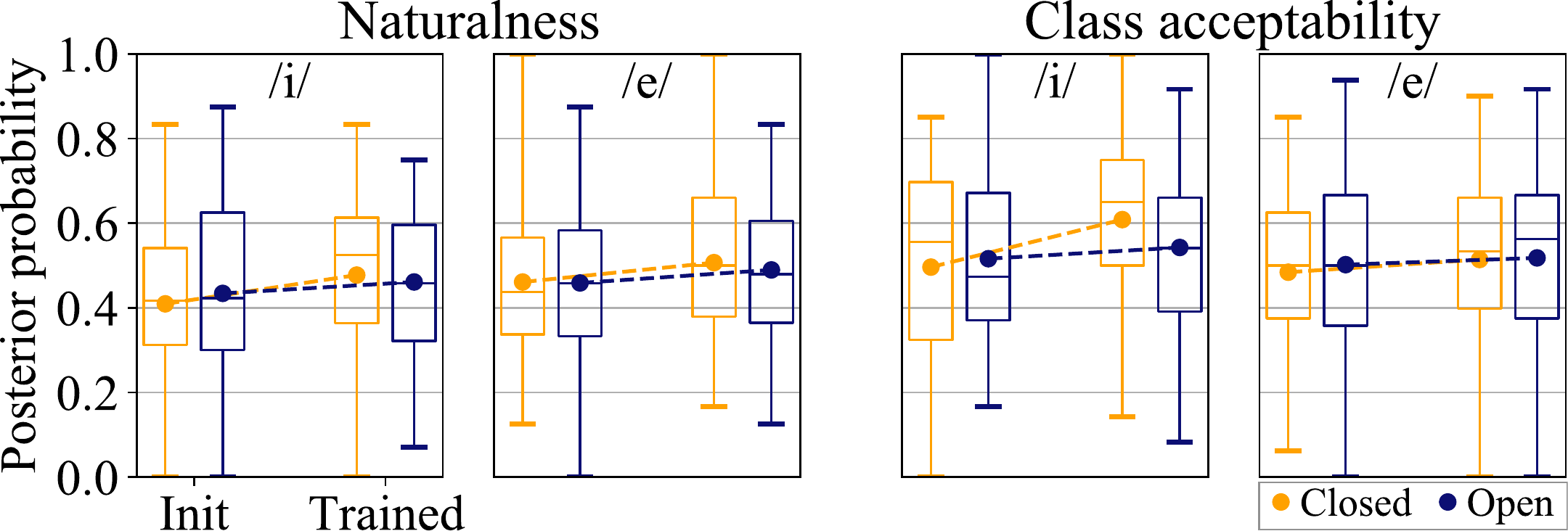}{Posterior probability of data generated from initialized (``Init'') or trained (``Trained'') generators. The boxes indicate first, second (i.e., median), and third quantiles. The line plot indicates mean value.}
        
    \subsection{Difference between real-data and human-acceptable distributions} \label{sec:diff}
        First, we confirmed that the human-acceptable distributions were wider than the real-data distributions. We carried out two tests: evaluations of humans' tolerance of naturalness and class acceptability. In this experiment, class acceptability denotes whether the data sounded like the presented phoneme. We split the two-dimensional space into grids and generated a speech waveform for every grid. Then, we presented a speech waveform to a listener, and the listener rated the naturalness on a 5-point scale from 1 (bad) to 5 (excellent). Next, we presented a speech waveform and a phoneme label (/i/ or /e/) to a listener, and the listener rated the class acceptability on the same scale. One-through-five of the obtained scores corresponded to $0.00, 0.25, 0.50, 0.75,$ and $1.00$ of the posterior probability ($D_{\mathrm S}\pt{\hatx_n}$ or $D_{\mathrm C}\pt{\hatx_n,\cc_n}$), respectively. We used the Lancers crowdsourcing platform~\cite{lancers} to execute the evaluations. The posterior probabilities were averaged for each grid. Each grid was scored by at least five listeners. The total number of listeners was 105.
        
        \Fig{apos} shows the results. As described in \Sec{expcond}, the real data ware normalized to have zero-mean and unit-variance, and the ACGAN represents this range. However, as shown in \Fig{apos}, the human-acceptable distributions, i.e., darker zones of the color maps, were wider than the real-data distribution for both naturalness and class acceptability. Therefore, we obtained support for the HumanACGAN being able to represent this distribution, which an ACGAN cannot adequately represent.
    
    \subsection{Transition of generated data during training} \label{sec:learnmodel}
        Next, we executed the HumanACGAN training and qualitatively evaluated the generated data and approximated gradient. The prior noise followed a two-dimensional uniform distribution $U(0, 1)$. We randomly generated this prior noise before the training and fixed it during the iterations. Half of the data belonged to the class label /i/, and the remaining data belonged to /e/. The generator was a small feed-forward neural network consisting of a two-unit input layer, $2\times4$-unit sigmoid hidden layers, and a two-unit linear output layer. The output layer was conditioned using the two dimensional one-hot class label vector. We performed random initialization of the generator until these conditions were satisfied: The data generated from the initialized generator should cover ranges of the higher posterior probabilities of naturalness and class acceptability, and the data of each class should leak into the other class distribution for determining the ability to classify. We empirically set the hyperparameter $\lambda = 2$ and used the gradient descent method with a learning rate of $\alpha = 0.0005$ for the training. We used Chainer~\cite{chainerproc} for the implementation. The number of generated data $N$, the number of perturbations $R$, the number of training iterations, and the standard deviation of NES $\sigma$ were set to $50$, $5$, $4$, and $2.0$, respectively. We carried out two perceptual evaluations during the HumanACGAN training. First, we presented two speech waveforms to a listener, and the listener rated which one was more natural on a 5-point scale: 1: the first one, 3: equal, and 5: the second one. Next, we presented two speech waveforms and one phoneme label (/i/ or /e/) to a listener, and the listener rated which one sounded like the presented phoneme on the same scale. One-through-five of the obtained scores corresponded to $1.0, 0.5, 0.0, -0.5,$ and $-1.0$ of the difference in the posterior probabilities ($\Delta D_{\rm S}\pt{\hatx_n}$ or $\Delta D_{\rm C}\pt{\hatx_n,\cc_n}$), respectively.
        
        \Fig{grads} shows the gradient of every data point. The posterior probability of \Fig{apos} is also drawn for reference, but note that we never used it during the training. We can say that the gradient by both naturalness and class acceptability points to each darker range. This qualitatively indicates that the gradient shown in \Eqs{grad_S}, \Eqss{grad_C} was properly estimated. \Fig{init_after} shows the data generated from initialized and trained generators. On the basis of posterior probabilities shown in \Fig{apos}, we can say that the training makes the data move so that they have higher posterior probabilities of both naturalness and class acceptability. This qualitatively indicates that the loss functions shown in \Eqs{L_S}, \Eqss{L_C} improved the generator to represent the conditional human-acceptable distributions.
        
    \subsection{Increase in posterior probabilities during training}
        Finally, we quantitatively verified that the training increases the posterior probabilities of naturalness $D_{\mathrm S}\pt{\cdot}$, and class acceptability $D_{\mathrm C}\pt{\cdot}$. We prepared two types of data: closed and open. The closed data were generated from the prior noise used while the training, and open data was generated from a newly sampled prior noise that was not used during the training. The posterior probabilities of the closed/open data were scored in the same manner as the ones in \Sec{diff}. The total number of listeners was 160.

        \Fig{boxplot} shows the box plots of the posterior probability. The training iteration increased the posterior probabilities of both naturalness and class acceptability with not only the closed data but also the open data. Therefore, we can say that our training can increase the objective values consisting of posterior probabilities and that the generator of the HumanACGAN can represent conditional human-acceptable distributions.
        
\vspace{-1mm} 
\section{Conclusion} \vspace{-2mm}
    We proposed the HumanACGAN, which can conditionally represent humans' perceptually acceptable distributions. We reconfigured a discriminator and an auxiliary classifier of an ACGAN to utilize humans' perceptual evaluations. The DNN-based conditional generator of the HumanACGAN was trained using two kinds of human's perceptual evaluations: naturalness and class acceptability, which were used for the discriminator and the auxiliary classifier, respectively. We evaluated the effectiveness of the HumanACGAN using qualitative and quantitative experiments. We are planning to expand the scalability of the HumanACGAN in terms of scalability to the data size and dimensionality as part of our future work.

\textbf{Acknowledgements:}
Part of this work was supported by the MIC/SCOPE \#182103104.


\bibliographystyle{IEEEbib}
\bibliography{tts}

\begin{thebibliography}{10}

\bibitem{goodfellow14gan}
I.~Goodfellow, J.~Pouget-Abadie, M.~Mirza, B.~Xu, D.~WardeFarley, S.~Ozair,
  A.~Courville, and Y.~Bengio,
\newblock ``Generative adversarial nets,''
\newblock in {\em Proc. NIPS}, Montreal, Canada, Dec. 2014, pp. 2672--2680.

\bibitem{kingma2013vae}
D.~Kingma and M.~Welling,
\newblock ``Auto-encoding variational bayes,''
\newblock {\em arXiv}, vol. abs/1312.6114, 2013.

\bibitem{dinh2014nice}
L.~Dinh, D.~Krueger, and Y.~Bengio,
\newblock ``{NICE}: Non-linear independent components estimation,''
\newblock in {\em Proc. ICLR}, San Diego, U.S.A., May 2015.

\bibitem{saito18advss}
Y.~Saito, S.~Takamichi, and H.~Saruwatari,
\newblock ``Statistical parametric speech synthesis incorporating generative
  adversarial networks,''
\newblock {\em IEEE/ACM Transactions on Audio, Speech, and Language
  Processing}, vol. 26, no. 1, pp. 84--96, Jan. 2018.

\bibitem{hono2019singing}
Y.~Hono, K.~Hashimoto, K.~Oura, Y.~Nankaku, and K.~Tokuda,
\newblock ``Singing voice synthesis based on generative adversarial networks,''
\newblock in {\em proc. ICASSP}, Brighton, United Kingdom, May 2019, pp.
  6955--6959.

\bibitem{fujii2020humangan}
K.~Fujii, Y.~Saito, S.~Takamichi, Y.~Baba, and H.~Saruwatari,
\newblock ``Human{GAN}: generative adversarial network with human-based
  discriminator and its evaluation in speech perception modeling,''
\newblock in {\em Proc. ICASSP}, Barcelona, Spain, May 2020, pp. 6239--6243.

\bibitem{sagisaka88}
Y.~Sagisaka,
\newblock ``Speech synthesis by rule using an optimal selection of non-uniform
  synthesis units,''
\newblock in {\em Proc. ICASSP}, New York, U.S.A., Apr. 1988, pp. 679--682.

\bibitem{stylianou88}
Y.~Stylianou, O.~Capp\'{e}, and E.~Moulines,
\newblock ``Continuous probabilistic transform for voice conversion,''
\newblock {\em IEEE Transactions on Speech and Audio Processing}, vol. 6, no.
  2, pp. 131--142, Mar. 1998.

\bibitem{mirza2014conditional}
M.~Mirza and S.~Osindero,
\newblock ``Conditional generative adversarial nets,''
\newblock {\em arXiv preprint arXiv:1411.1784}, 2014.

\bibitem{odena2017conditional}
A.~Odena, C.~Olah, and J.~Shlens,
\newblock ``Conditional image synthesis with auxiliary classifier {GAN}s,''
\newblock in {\em Proc. ICLR}, Vancouver, Canada, Apr. 2018.

\bibitem{chiu2020human}
C.~Chiu, Y.~Koyama, Y.~Lai, T.~Igarashi, and Y.~Yue,
\newblock ``Human-in-the-loop differential subspace search in high-dimensional
  latent space,''
\newblock {\em ACM Transactions on Graphics (TOG)}, vol. 39, no. 4, pp. 85--1,
  2020.

\bibitem{peterson2018capturing}
J.~Peterson, J.~Suchow, K.~Aghi, A.~Ku, and T.~Griffiths,
\newblock ``Capturing human category representations by sampling in deep
  feature spaces,''
\newblock {\em arXiv preprint arXiv:1805.07644}, May 2018.

\bibitem{ilyas18blackboxlimitedquery}
A.~Ilyas, L.~Engstrom, A.~Athalye, and J.~Lin,
\newblock ``Black-box adversarial attacks with limited queries and
  information,''
\newblock in {\em Proc. ICML}, Stockholm, Sweden, Jul. 2018, vol.~2, pp.
  2137--2146.

\bibitem{goodfellow16gantutorial}
I.~Goodfellow,
\newblock ``{NIPS} 2016 tutorial: Generative adversarial networks,''
\newblock in {\em Proc. NIPS}, Barcelona, Spain, Dec. 2016.

\bibitem{jvpd_corpus}
``Vowel database: Five {J}apanese vowels of males, females, and children along
  with relevant physical data ({JVPD}),''
  \url{http://research.nii.ac.jp/src/en/JVPD.html}.

\bibitem{morise16world}
M.~Morise, F.~Yokomori, and K.~Ozawa,
\newblock ``{WORLD}: a vocoder-based high-quality speech synthesis system for
  real-time applications,''
\newblock {\em IEICE transactions on information and systems}, vol. E99-D, no.
  7, pp. 1877--1884, Jul. 2016.

\bibitem{morise16d4c}
M.~Morise,
\newblock ``{D4C}, a band-aperiodicity estimator for high-quality speech
  synthesis,''
\newblock {\em Speech Communication}, vol. 84, pp. 57--65, Nov. 2016.

\bibitem{lee01julius}
A.~Lee, T.~Kawahara, and K.~Shikano,
\newblock ``Julius --- an open source real-time large vocabulary recognition
  engine,''
\newblock in {\em Proc. EUROSPEECH}, Aalborg, Denmark, Sep. 2001, pp.
  1691--1694.

\bibitem{lancers}
``Lancers,'' \url{https://www.lancers.jp/}.

\bibitem{chainerproc}
S.~Tokui, R.~Okuta, T.~Akiba, Y.~Niitani, T.~Ogawa, S.~Saito, S.~Suzuki,
  K.~Uenishi, B.~Vogel, and H.~Y Vincent,
\newblock ``Chainer: A deep learning framework for accelerating the research
  cycle,''
\newblock in {\em Proc. KDD}, Anchorage, U.S.A., Aug. 2019, pp. 2002--2011.

\end{thebibliography}

\end{document}